\documentclass{aa}
\def\e{et~al.\ }
\def\es{{\rm erg\,s^{-1}} }
\def\r{R_{\rm m}}
\newcommand{\be}{\begin{equation}}
\newcommand{\ee}{\end{equation}}
\newcommand{\bdm}{\begin{displaymath}}
\newcommand{\edm}{\end{displaymath}}

\begin{document}

    \title{On the accretion flow geometry in A0535+26}

   \author{Nazar R. Ikhsanov\inst{1,2}, Valeri M. Larionov\inst{3,4}
and Nina G. Beskrovnaya\inst{2}}

 \offprints{N.R.~Ikhsanov \\ \email{ikhsanov@mpifr-bonn.mpg.de}}

   \institute{Max-Planck-Institut f\"ur Radioastronomie, Auf dem
              H\"ugel 69, D-53121 Bonn, Germany
              \and
              Central Astronomical Observatory of the Russian Academy of               Science at Pulkovo,
              Pulkovo 65--1, 196140 Saint-Petersburg, Russia
              \and
              Astronomical institute of St.\,Petersburg University,
              St.\,Petersburg, Russia
              \and
             Isaac Newton Institute of Chile, St.Petersburg Branch.
}

   \date{Received 28 November 2000 / Accepted 20 March 2001}

\authorrunning{Ikhsanov N.R. \e}

    \abstract{
The geometry of accretion flow in the Be/X-ray transient A0535+26 is
explored. It is shown that neither moderate nor giant X-ray flaring
events observed in the system can be interpreted within the
spherically symmetrical accretion model and hence the formation of
an accretion disk around the neutron star magnetosphere during the
both types of flares is required. The accretion disk can be formed
at the periastron if (i)\,the expansion velocity of the
Be star envelope in the equatorial plane is $V_{\rm wr} < 150\,{\rm
km\,s^{-1}}$ and (ii)\,the parameter  accounting for the accretion
flow inhomogeneities, $\xi$, satisfies the following condition:
$\xi \ga 0.16\ \dot{M}_{17}^{-1/7}$, where $\dot{M}_{17}$
is the rate of mass capture by the neutron star expressed in units
of $10^{17}\,{\rm g\,s^{-1}}$.
We suggest that the `missing' outburst phenomenon can be associated
with the spherically symmetrical accretion onto the interchange-stable
magnetosphere of the neutron star. The average spin up rate
of the neutron star during moderate flares $\la 3.5 \times
10^{-12}\,{\rm Hz\,s^{-1}}$ is predicted.
}

   \maketitle


   \section{Introduction}

A0535+26 is a close binary system which contains the O9.7\,IIe --
B0\,IIIe star HDE~245770 and a rotating, strongly magnetized neutron
star. The optical companion is the source of intensive
stellar wind ($\sim 10^{-8} {\rm M_{\sun}\,yr^{-1}}$), which
consists of a high velocity, low density component at high latitudes
and a low velocity high density circumstellar disk at the equatorial
plane. The star is observed to be variable in the optical-IR on time
scales from a few months to a few years (Giovannelli \& Graziati
\cite{gg92}; Haigh \e \cite{haigh99}).

A0535+26 was discovered as a source of hard X-ray emission by
{\it Ariel}~V during the giant outburst in 1975 (Rosenberg \e \cite{r75};
Coe \e \cite{c75}). Further observations have shown this system to be a
transient X-ray source in which the neutron star (a 103\,s
pulsar with the surface magnetic field of $B\approx 10^{13}$\,G: Kanno
\cite{kanno80}; Dal Fiume \e \cite{df88}) is in an eccentric orbit
($e= 0.47\pm0.02$) of the $P_{\rm orb} \simeq 110.3\pm0.3$\,days period
around the Be companion (Finger \e \cite{f96}).
On the time scale of $P_{\rm orb}$ the system may exhibit either no outburst
(`missing' outburst), a moderate (`normal') or a giant outburst.
Almost all moderate X-ray outbursts occur around a certain orbital phase.
The X-ray flux at the maximum of moderate flares lies within the interval
$0.1\div 0.8$\,Crab (in the 2--10\,keV band) and their duration is
$\sim 10\div 15$\,days. Giant X-ray flares are of longer duration
(up to 40\,days) and the maximum X-ray flux detected during these events
exceeds that of moderate flares by almost an order of magnitude.
Some of giant flares\footnote{Since 1975 five giant flares have been
observed in the system} are delayed in phase with respect to the moderate
flares by 10--15\,days. The spin up behaviour of the neutron star and
the quasi periodic oscillations (QPOs) have been reported for the
giant flare 1994 (Finger \e \cite{f96} and references therein).

The commonly accepted interpretation of the X-ray transient behaviour
of A0535+26 is based on the so called eccentric orbit model. According to
this model the rate of mass capture by the neutron star from the
stellar wind of the normal companion depends on the orbital phase.
It reaches its maximum value during the periastron passage when the
neutron star moves through the dense equatorial disk of the Be star.
As a result, all X-ray flares are expected to occur at the same
orbital phase, which is associated with the periastron, $\phi=0.0$
(see Giovannelli \& Graziati \cite{gg92}).

Some effort has been made to explain the difference in observed
properties of the moderate and giant flares. De~Loore \e
(\cite{del84}) suggested that giant flares occur due to
temporal enhancement of plasma density in the vicinity of the
neutron star due to the Be star envelope ejection.  
Another possibility has been discussed by Motch \e (\cite{m91})
who pointed out that moderate and giant flares can be triggered 
by different accretion mechanisms: spherical and disk accretion, 
respectively. Finally, Negueruela \e (\cite{neg98}) associated 
giant flares with large-scale perturbations in the tidally 
truncated circumstellar disk of the Be companion.

The phenomenon of `missing' flares remains one of the most puzzling points
in the interpretation of A0535+26. First it was assumed that the absence of
X-ray outburst at the orbital phase $\phi=0.0$ is caused by the lower
Be star activity and, correspondingly, the decrease of plasma
density in the stellar envelope (Giovannelli \& Graziati \cite{gg92}).
However, systematic multiwavelength studies of the system reported
recently (Clark \e \cite{cts98} and references therein) did not
confirm this hypothesis:  `missing' flare phenomena occured during
the period when the optical--IR flux of the star was increasing.

In this paper we explore a possibility to explain the observed diversity of
X-ray flaring in A0535+26 in terms of variable geometry of accretion flow
onto the magnetosphere of the neutron star.
In the next section the spherically symmetrical accretion onto the
neutron star magnetosphere is discussed. We find that in
the particular case of A0535+26 the magnetospheric boundary is stable
with respect to interchange instabilities. The upper limit to
X-ray luminosity in this case is two orders of magnitude smaller than
the X-ray luminosity observed during the moderate flares. On this
basis, we suggest that the accretion flow during the X-ray flares in
the system has a disk-like geometry. The condition for the accretion
disk formation is discussed in section~3. Application of our approach
to the interpretation of the `missing' outburst phenomenon is the
subject of section~4. In section~5 we examine possible observational
signs of an accretion disk during moderate flares. The results are
summarized in section~6.

    \section{Spherically symmetrical accretion}

Within the spherically symmetrical accretion model, the plasma
captured by the neutron star from the wind of the Be companion is
assumed to flow toward the compact star in an almost radial direction
with the free-fall velocity,
    \bdm
 V_{\rm ff}(R) = \sqrt{2 G M_{\rm ns}/R}.
    \edm
Interaction of the accreting plasma with the magnetic field of the
neutron star leads to the formation of a magnetosphere that, to a
first approximation, prevents the plasma from reaching the star
surface. In this context, the mode by which the accreting plasma
enters the magnetosphere proves to be a key question in the
interpretation of the accretion-powered sources.

As shown in a previous paper (Ikhsanov \cite{i00})
the rate of plasma penetration into the magnetosphere due to diffusion
process,
  \be\label{maxdifrad}
\dot{M}_{\rm diff} \sim 10^{12}\,{\rm g\,s^{-1}} \zeta^{1/2}
\mu_{31}^{-1/14} M_{1.5}^{1/7}
\left(\frac{\dot{M}_{\rm c}}{10^{17}\,{\rm g\,s^{-1}}}\right)^{11/14}
  \ee
is too small to explain the observed X-ray luminosity of the system.
Here, $\mu_{31}$ and $M_{1.5}$ are the magnetic moment and mass of
the neutron star expressed in units of $10^{31}\,{\rm G\,cm^3}$ and
$1.5 M_{\sun}$, respectively, $\zeta$ is the efficiency of the
diffusion process ($\zeta < 1$) and $\dot{M}_{\rm c}$ is the mass
of the surrounding material with which the neutron star moving
through the wind of the Be companion interacts in a time unit (see
section~4).

If the penetration process is governed by the reconnection
of the field lines, the mass accretion rate onto the surface of the
neutron star in A0535+26 can be evaluated as follows
    \be\label{dotmrec}
\dot{M}_{\rm rec} \simeq 10^{15}\,{\rm g\,s^{-1}}\
\left[\frac{\alpha_{\rm R}}{0.1}\right]
\left[\frac{\lambda_{\rm m}}{0.1 \r}\right]
\left(\frac{\dot{M}_{\rm c}}{10^{17}\,{\rm g\,s^{-1}}}\right),
  \ee
where $\alpha_{\rm R}$ is the efficiency of the reconnection process
and $\lambda_{\rm m}$ is the characteristic scale of the magnetic
field in the accretion flow.
As has been recently shown by Ikhsanov (\cite{i00}), the
reconnection-driven accretion model allows one to interpret the
quiescent X-ray emission observed from A0535+26. At the same time,
for the same accretion mechanism to be responsible for the system
X-ray emission during flares, the number density of plasma in the
inner radius of the Be star circumstellar disk should be in excess of
$\sim 10^{13}\,{\rm cm^{-1}}$. This value, however, is rather
non-realistic since it exceeds  by an order of magnitude the number
density of plasma in the disk evaluated by Clark \e (\cite{cscr98}).

The rate of plasma entry into the magnetosphere could be 
higher if the magnetospheric boundary is unstable with respect to
interchange instabilities. However, the application of this scenario
to the particular case of A0535+26 encounters a serious problem.

Arons \& Lea (\cite{al76}) and Elsner \& Lamb (\cite{el76}) have
shown that the magnetospheric boundary of a neutron star
undergoing spherical accretion is convex towards the accreting
plasma. That is why the gradient of the field is such as to
stabilize the boundary. In this situation, the magnetosphere is
interchange unstable if the effective gravitational acceleration
at the  boundary has a positive sign:
     \be\label{eff}
g_{\rm eff}\ = \frac{G M_{\rm ns}}{\r^{2}(\kappa)} \cos{\kappa} -
\frac{V_{\rm T_{\rm i}}^{2}(\r)}{R_{\rm c}(\kappa)} > 0,
      \ee
where $R_{\rm c}$ is the curvature radius of the field lines,
$\kappa$ is the angle between the radius vector and the outward
normal to the magnetospheric boundary and $V_{\rm T_{\rm i}}(\r)$ is
the ion thermal velocity of the accreting plasma at the boundary.

This condition can be satisfied if the cooling time of plasma at the
boundary is smaller than the free-fall time, i.e. a typical time of
plasma heating due to accretion. According to Arons \& Lea and
Elsner \& Lamb this can be realized only if the luminosity
of the X-ray source is
\be
\hspace{0.5cm}L_{\rm x} \ga L_{\rm cr} \simeq 6.5 \times 10^{36}\ \es\ \times
\ee
\bdm
\hspace{1.5cm}\times \left[\frac{\mu}{10^{31} {\rm G}}\right]^{1/4}\
\left[\frac{M_{\rm ns}}{1.5 M_{\sun}}\right]^{1/2}
\left[\frac{R_{\rm ns}}{10^6 {\rm cm}}\right]^{-1/8}
\edm
and thus, the Compton cooling is effective. Other mechanisms, like 
free-free and/or cyclotron cooling, are not effective in this
case because of the relatively low density of the accretion flow
beyond the magnetospheric boundary and the insufficiently high
strength of the magnetic field at the Alfv\'en surface, respectively.

The condition $L_{\rm x} \ga L_{\rm cr}$ in the case of A0535+26
is satisfied only during giant flares and at the maximum of a few
moderate flares. Excluding these events, the instabilities of the
boundary are suppressed and the rate of plasma penetration into the
magnetic field of the neutron star is limited by $\dot{M}_{\rm rec}$.
This makes the interpretation of flares with luminosity
$L_{\rm x} < 6\times 10^{36}\es$ in the frame of the spherical
accretion approach very problematic.
Furthermore, for the effective penetration of plasma into the
star magnetosphere, the luminosity of the X-ray source should already
be above the critical value $L_{\rm cr}$.
This situation could be realized if an additional process operates
which leads to periodical ($\sim 111$\,days) outbursts with
luminosity  $L_{\rm x}\ga 7\times 10^{36} \es$ inside the neutron star
magnetosphere. However, extensive observations of A0535+26 have shown
no evidence for such a process. On this basis, we focus on the
investigation of another possibility. Here we explore whether the
observed transient X-ray flaring in A0535+26 can be explained in
terms of deviations of accretion flow geometry from the spherically
symmetric case.

	     \section{Transient disk formation}

The problem of accreting plasma penetration into the magnetic field
of the neutron star described in the previous section can be avoided
if the accretion disk forms around the magnetosphere. As 
shown by Ghosh \& Lamb (\cite{gl79}), the rate of plasma penetration
from the accretion disk into the magnetosphere of a neutron star
is large enough to power the X-ray emission of the brightest X-ray
pulsars. In this context the question about the effective penetration
of the accreting plasma into the neutron star magnetosphere is
reduced to the question about the conditions for accretion disk
formation in the system.

    	 \subsection{Wind-fed mass transfer}

The Roche lobe radius of the Be component can be estimated following
Eggleton (\cite{e83}), taking the mass ratio
$q=M_{\rm Be}/M_{\rm ns} =10$ as
  \be\label{roche}
R_{\rm L_{\rm Be}} \simeq 3.4 \times 10^{12}\,{\rm cm}\
\left(\frac{a_0}{8.6 \times 10^{12}\,{\rm cm}}\right),
   \ee
where $a_0$ is the orbital separation at the periastron\footnote{Here
we assume the eccentricity of the orbit to  be $e=0.5$.}. Hence, the
ratio of the radius of the optical companion  to its mean Roche lobe
radius in A0535+26 is
   \bdm
\frac{R_{\rm Be}}{R_{\rm L_{\rm Be}}} \la 0.3
\left(\frac{R_{\rm Be}}{14 R_{\sun}}\right)
\left(\frac{R_{\rm L_{\rm Be}}}{3.4 \times 10^{12}\,{\rm
cm}}\right)^{-1}.   \edm
This clearly indicates that the optical component underfills its Roche
lobe and thus, the wind-fed mass transfer is realized in the system.

	\subsection{Condition for an accretion disk formation}

For a disk to form, the specific angular momentum of matter captured
by the compact star must be sufficient to allow it to enter a Keplerian
circular orbit around the magnetosphere:
   \be\label{dr}
R_{\rm m} < R_{\rm circ} \simeq
\frac{\dot{J}^{2}}{\dot{M}_{\rm c}^{2} G M_{\rm ns}},
   \ee
where $R_{\rm circ}$ is the circularization radius and $\dot{J}$
is the rate of accretion of angular momentum, which is
   \be\label{aar}
\dot{J}=\xi\dot{J}_{0} =\xi \left(\frac{1}{2} \Omega_{\rm orb}
r_0^{2} \dot{M}_{\rm c}\right).
   \ee
Here, $\Omega_{\rm orb}$ and $r_0$ are the orbital angular velocity
and the accretion radius of the neutron star, respectively.
Parameter $\xi$ is the factor by
which the average rate of accretion of angular momentum is reduced
due to inhomogeneities (the velocity and density gradients) in the
accretion flow.

If a neutron star is a component of a close binary system, its
accretion radius is
  \be\label{accrr}
r_0\simeq\left\{
\begin{array}{lc}
r_{\alpha}=2GM_{\rm ns}/V_{\rm rel}^2, \hspace{1mm}& {\rm for} \hspace{5mm}
V_{\rm rel} > V_0, \\
& \\
R_{\rm L_{\rm ns}}, \hspace{1mm}& {\rm for} \hspace{5mm} V_{\rm rel}
\la V_0,
 \end{array}
  \right.
   \ee
where $R_{\rm L_{\rm ns}}$ is the neutron star Roche lobe radius and
the velocity $V_0$ is determined as $r_{\alpha}(V_0)=R_{\rm
L_{\rm ns}}$. Using the parameters of A0535+26, we find at the
periastron point    \be\label{rnsr}
R_{\rm L_{\rm ns}}\simeq  1.8\times 10^{12}\ {\rm cm}
\left(\frac{a_0}{8.6 \times 10^{12}\,{\rm cm}}\right),
  \ee
and the value of $V_0$:
  \be\label{vnul}
V_0 = 150\,{\rm km\,s^{-1}}
\left[\frac{M_{\rm ns}}{1.5 M_{\sun}}\right]^{1/2}
\left[\frac{a_0}{8.6\times 10^{12}\ {\rm cm}}\right]^{-1/2}.
   \ee

The parameter $\xi$ has been evaluated from the numerical simulations
of mass transfer in wind-fed close binaries (e.g. Anzer \e
\cite{anzer87}; Taam \& Fryxell \cite{taam88}; Matsuda \e
\cite{matsuda91}). These studies revealed the parameter $\xi$ to be
variable, due to flip-flop instability of the accretion flow on a time
scale of a few\,$\times r_0/V_{\rm ff}(r_0)$ with an average value
$\bar{\xi}=0.2$. Combining this result with Eqs.~(\ref{dr} --
\ref{vnul}) we find the conditions for the accretion disk formation
at the periastron in A0535+26 in the following form
  \be\label{ksi1}
\left\{
\begin{array}{ll}
V_{\rm rel} < 160\,{\rm km\,s^{-1}}\ \xi_{0.2}^{1/4}\
\dot{M}_{17}^{1/28}  & {\rm for}
\hspace{5mm} V_{\rm rel} > V_0, \\
 & \\
\xi\ \ga\  0.16\ \dot{M}_{17}^{-1/7} & {\rm for} \hspace{5mm} V_{\rm
rel} \la V_0.
   \end{array}
    \right.
     \ee

On this basis the following conclusions on the flaring scenario in
A0535+26 can be made.

  \section{`Missing' flare phenomenon}

The average mass capture rate by the neutron star at the periastron in
A0535+26 can be evaluated as follows
   \be\label{mc1}
\dot{M}_{\rm c}(a_0) = \pi r_0^2(a_0)\ \rho(a_0)\ V_{\rm rel}(a_0),
   \ee
where $\rho(a_0)$ is the density of plasma surrounding the neutron star
at the periastron.

Following Lamers \& Waters (\cite{lw87}), the density and velocity in
the circumstellar disk can be expressed as
   \be\label{rho0v0}
\left[
  \begin{array}{l}
\rho(R) = \rho_0 (R/R_{\rm Be})^{-n},\\
\\
V_{\rm wr}(R) = V_{\rm r0} (R/R_{\rm Be})^{n-2},
   \end{array}
\right.
   \ee
where $\rho_0$ and $V_{\rm r0}$ are the density and velocity of plasma
at the inner disk radius and the value of $n$ lies within the interval
$2.3 \div 3.3$. For the case of A0535+26, we set hereafter $n=3$ (for
a discussion see Clark \e \cite{clark99}).

Investigating the properties of Paschen emission lines in A0535+26,
Clark \e (\cite{cscr98}) have estimated the plasma density in the
circumstellar disk to be $10^{12}\,{\rm cm^{-3}}$. Assuming these
lines to be generated in the inner part of the disk one finds $\rho_0
= (1\div 2)\times 10^{-12}\,{\rm g\,cm^{-3}}$.

The value of $V_{\rm r0}$ suggested by Lamers \& Waters (\cite{lw87})
for field Be stars lies within the interval $2-20\,{\rm km\,s^{-1}}$.
Applying this to A0535+26 we find the mass outflow rate in the disk
of $(1\div 10) \times 10^{-8}\,{\rm M_{\sun}\,yr^{-1}}$, which is in
good agreement with that estimated by
Clark \e (\cite{clark99})\footnote{Here we adopted the disk opening
angle of $\theta=15^{\degr}$.}.

Finally, the relative velocity can be presented in the following form
    \be\label{vrel1}
\vec{V}_{\rm rel} = \vec{V}_{\rm ns} + \vec{V}_{\rm w},
   \ee
where $\vec{V}_{\rm ns}$ is the linear velocity of the neutron star
orbiting around the Be companion and $\vec{V}_{\rm w}$ is the stellar
wind velocity in the frame of the Be star:
    \be\label{vw}
\vec{V}_{\rm w} = V_{\rm wr} \vec{e}_{\rm r} + V_{\rm w\phi}
\vec{e}_{\phi}.
    \ee
Here $\vec{e}_{\rm r}$ and $\vec{e}_{\phi}$ are the unit vectors in
the radial and azimuthal directions, respectively. At the periastron,
one can set\footnote{Here we take into account that $M_{\rm ns}
\ll M_{\rm Be}$.}
   \be
V_{\rm ns}(a_0) = \frac{2 \pi \bar{a}}{P_{\rm orb}}
\sqrt{\frac{1+e}{1-e}},
   \ee
   \be\label{vwphi}
V_{\rm w\phi}(a_0) \la \sqrt{GM_{\rm Be}/a_0},
   \ee
where $\bar{a}$ is the average value of the orbital separation.

Thus, using these values in Eq.~(\ref{mc1}) and calculating $r_0$
from Eq.~(\ref{accrr}) one can express the mass capture rate by the
neutron star during the periastron passage as
   \be\label{dotma0}
\dot{M}_{\rm c}(a_0)\mid_{(V_{\rm rel} \ga V_0)}\ \simeq\ 10^{17}\,{\rm
g\,s^{-1}}\ M_{1.5}^2\  \times
   \ee
    \bdm
\hspace{1.5cm}\times
\left(\frac{N(a_0)}{10^{9}\,{\rm cm^{-3}}}\right)
\left(\frac{V_{\rm rel}(a_0)}{180\,{\rm km\,s^{-1}}}\right)^{-3},
    \edm
and
  \be
\dot{M}_{\rm c}(a_0)\mid_{(V_{\rm rel} < V_0)}\ \simeq\ 7
\times 10^{16}\,{\rm g\,s^{-1}}\ M_{1.5}^2\  \times
   \ee
    \bdm
\hspace{1.5cm}\times
\left(\frac{N(a_0)}{10^{9}\,{\rm cm^{-3}}}\right)
\left(\frac{V_{\rm rel}(a_0)}{40\,{\rm km\,s^{-1}}}\right).
    \edm

On the other hand, the rate of mass accretion onto the neutron star
surface during the missing flare
phenomenon is at least an order of magnitude smaller than the mass
capture rate evaluated above. Hence, if the absence of a X-ray
outburst at the periastron is caused by the lower activity of the Be
star, the density of the circumstellar disk during this time should be
smaller than its average value by at least a factor of 10. Under
this condition, the disk is expected to be faint or almost invisible.

However, systematic optical-IR observations have shown the `missing'
flare phenomena to occur also during the periods when the
circumstellar disk is in its brightest state (see Clark \e
\cite{cts98}). This suggests that the low activity of the Be star is
not the only reason for the `missing' flare phenomenon and an
additional factor, which leads to suppression of X-ray flaring at the
periastron, operates in the system.

Within the approach presented in this paper, the `missing' flare
phenomenon can be associated with the spherical geometry of the
accretion
flow and, correspondingly, the low rate of plasma penetration into the
magnetosphere of the neutron star. According to condition~(\ref{ksi1})
this situation is realized if either the relative velocity exceeds the
critical value or if the parameter $\xi$ is small enough to prevent
the formation of the accretion disk around the neutron star
magnetosphere.

Combining Eqs.~(\ref{vrel1}--\ref{vwphi}) we find that an accretion
disk around the neutron star cannot form if the radial velocity
of plasma in the circumstellar disk of the Be companion is
$V_{\rm wr}(a_0) \ga 150\,{\rm km\,s^{-1}}$. Under this condition
the mass accretion rate onto the neutron star surface is limited
by $\dot{M}_{\rm rec}$ and thus, the X-ray luminosity of the neutron
star during the periastron passage remains almost at the quiescent
level.

In the case of $V_{\rm wr}(a_0) < 150\,{\rm km\,s^{-1}}$, the missing
flare phenomenon can occur only if the second part of 
condition~(\ref{ksi1}) is not satisfied. Analysis of this situation
is especially important if the Be companion of A0535+26 is
surrounded by the viscous decretion disk (Lee \e \cite{los91};
Porter \cite{p99}).
In this case the radial plasma velocity at the periastron is
expected to be relatively small, i.e. only a
few\,$\times {\rm km\,s^{-1}}$ (Hanuschik \cite{h00};
Okazaki \cite{o00}). In this situation Eq.~(\ref{vw}) can be
simplified, neglecting the term $V_{\rm wr}$ that gives the value of
the relative velocity at the periastron,
$V_{\rm rel}(a_0)\simeq 4\,10^6\,{\rm cm\,s^{-1}}$ and,
correspondingly, the strength of the stellar wind
$\dot{M}_{\rm c}(a_0) \simeq\ 7\,10^{16}\,{\rm g\,s^{-1}}$.
Under these conditions the formation of the accretion disk around the
neutron star in A0535+26 does not occur if $\xi \la 0.17$, i.e.
if the parameter $\xi$ is slightly smaller than its average value
$\bar{\xi}$. According to Taam \& Fryxell (\cite{taam88}), the
variations of the parameter $\xi$ in the wind-fed mass-exchange
close binaries is not unusual. In particular, they found the value
of $\xi$ to vary due to flip-flop instability within the interval
$0.05 \la \xi < 1$. The characteristic time of this instability in
the case of A0535+26 is
 \be
t_{\rm var}\simeq {\rm a~few} \times
\frac{r_0^{3/2}}{\sqrt{2GM_{\rm ns}}} \sim 4 \div 7\,{\rm days}.
   \ee
Since $t_{\rm var}$ exceeds the time interval in which the moderate
flares are scattered around the zero orbital phase ($\Delta
\phi\approx 0.02 \sim 2$\,days) one can envisage a situation in which
the value of $\xi$ at the periastron proves to be small enough for
the condition~(\ref{ksi1}) not to be satisfied. If this happens, the
geometry of the accretion flow is spherically symmetrical and
the mass accretion rate onto the neutron star surface is limited
by $\dot{M}_{\rm rec}$.

	 \section{Possible signs of accretion disk during moderate
flares}

The last question we briefly address in this paper is the
observational appearance of the accretion disk in A0535+26 during
moderate flares.

The conclusion about the presence of an accretion disk in the system
during giant flares has been made on the basis of X-ray observations
of the neutron star spin up behaviour and the QPOs (see Finger \e
\cite{f96} and references therein). Can these criteria be used to
conclude the presence or absence of the accretion disk during
moderate flares\,?

A neutron star undergoing disk accretion is spinning up if
   \be\label{pdot}
\frac{dI\omega}{dt}=K_{\rm su}-K_{\rm sd} > 0,
   \ee
where $I$ is the moment of inertia of the neutron star,
$\omega=2 \pi/P_{\rm s}$ and
$K_{\rm su}$ and $K_{\rm sd}$ are the accelerating and
decelerating torques applied to the neutron star, respectively.

Following Lipunov (\cite{l92}) we take
	\be\label{ksusd}
\begin{array}{l}
K_{\rm su} \approx \dot{M}_{\rm c}\  \sqrt{GM_{\rm ns}\r} \\
\\
K_{\rm sd} \approx k_{\rm t} \mu^2/R_{\rm cor}^3,
\end{array}
   \ee
where $k_{\rm t}$ is the dimensionless parameter of the order of unity
and $R_{\rm cor}$ is the corotational radius, which in the case of the
neutron star in A0535+26 is
   \be
R_{\rm cor} \simeq 3.8 \times 10^{9}\  P_{103}^{2/3}\
\left[\frac{M_{\rm ns}}{1.5 M_{\sun}}\right]^{1/3}
{\rm cm}.
   \ee

Putting Eq.~(\ref{ksusd}) into (\ref{pdot}) we find that the neutron
star in A0535+26 is spinning up if the X-ray luminosity in a flare is
   \be
L_{\rm x} \ga 5 \times 10^{35}\ \mu_{31}^2\ R_6\ P_{103}^{-7/3}
\left[\frac{M_{\rm ns}}{1.5 M_{\sun}}\right]^{-2/3}\ \es.
   \ee
This indicates that, in the frame of our approach, the neutron star
is expected to be spinning up in both the giant and moderate
flares. The spin up rate is $\dot{P} \propto L_{\rm x}^{6/7}$.
Hence, taking as the basis the spin up observations of the flare 1994
(Finger \e \cite{f96}), we can predict the spin up rate during
moderate flares to be
   \bdm
\dot{\nu}_{\rm fm} \simeq 3.5 \times 10^{-12}\,{\rm Hz\,s^{-1}}
\left[\frac{F_{\rm (2-10\,keV)}}{0.5 {\rm Crab}}\right]^{6/7},
  \edm
where $F_{\rm (2-10\,keV)}$ is the X-ray flux
during a moderate flare in the 2--10\,keV band.

The question whether it is possible to observe QPOs during moderate
flares is more complicated. The answer depends on whether the
accretion disk forming during a particular event is optically thick
or thin. The detailed investigation of this question is beyond the
scope of the present paper. Here we would like only to note that the
modeling of a developed accretion disk during moderate flares
faces some problems. In particular, assuming all material stored
in the disk to be accreted onto the neutron star surface one can
evaluate the mass of the disk as
   \bdm
M_{\rm d} \simeq 5 \times 10^{20}\,{\rm g}\ M_{1.5}^{-1}\ R_6
\left[\frac{E_{\rm f}}{10^{41}\,{\rm erg}}\right],
   \edm
where $E_{\rm f}$ is the total amount of energy released during the
flare. This, however, is essentially smaller than the average mass of
developed accretion disks estimated from observations of other close
X-ray binaries (see Lipunov \cite{l92}).
Furthermore, within the same assumption, the duration of a flare
depends on the radial velocity of plasma in the disk. This means that
the flare duration can be limited as
  \be
\Delta t_{\rm f} > \frac{R_{\rm out}}{V_{\rm dr}} =
\frac{m_{\rm p} \sqrt{GM_{\rm ns} R_{\rm out}}}
{\alpha k T(R_{\rm out})}.
  \ee
Here $T_{\rm d}(R_{\rm out})$ is the plasma temperature at the
outer radius of the disk, $R_{\rm out}$, and $V_{\rm dr}$ is the
radial velocity of plasma in the accretion disk which according to
Shakura \& Sunyaev (\cite{shs}) is
      \bdm
V_{\rm dr}\simeq \alpha c_{\rm s}^2/V_{\rm k},
      \edm
where $c_{\rm s}$ is the sound speed and $V_{\rm k}$ is the
keplerian velocity. If the disk is optically thick, i.e.
    \be\label{topt}
T_{\rm d}(R_{\rm out})= \left(\frac{\dot{M}_{\rm c} GM_{\rm ns}}
{4 \pi \sigma_{\rm sb} R_{\rm out}^3}\right)^{1/4},
    \ee
the flare duration is
   \be
\Delta t_{\rm f} > 50\,{\rm d}\ \alpha_{0.1}^{-1}\ M_{1.5}^{1/14}\
\mu_{31}^{5/7} \dot{M}_{18}^{-17/28}
\left[\frac{R_{\rm out}}{10 \r}\right]^{5/4}.
   \ee
The derived value is comparable with the duration of giant flares
but essentially exceeds the typical duration of moderate flares.
This demonstrates that if the accretion disk is formed during a
moderate flare, its structure is somewhat different from that of
developed optically thick disks and, therefore, the currently
suggested models of QPOs (e.g. Alpar \& Shaham \cite{as85}; Lamb
\e \cite{lamb85}; van der Klis \e \cite{vdk87}) cannot be applied
without additional investigation.

      \section{Conclusion}

Our main conclusion is that the accretion disk around the neutron
star magnetosphere forms during both the moderate and giant X-ray
flares in A0535+26. The grounds for this conclusion are the low rate
of plasma penetration into the neutron star magnetic field during the
spherically symmetrical accretion state that does not allow us to
explain the observed X-ray luminosity of the system during outbursts.
The low rate of plasma penetration into the magnetosphere is the
result of the stability of the magnetospheric boundary with respect to
interchange instabilities. The X-ray luminosity of the system within
the reconnection-driven accretion model is comparable with the
X-ray luminosity of A0535+26 observed the during quiescent state (for
discussion see Ikhsanov \cite{i00}).

The formation of an accretion disk around the magnetosphere of the
neutron star in A0535+26 can be expected during the periastron
passage if the following conditions are satisfied: (i) the expansion
velocity of the Be star envelope in the equatorial plane is $V_{\rm
wr} < 150\,{\rm km\,s^{-1}}$ and (ii) the parameter accounting for
the accretion flow inhomogeneities (density and velocity gradients) is
$\xi \ga 0.16\ \dot{M}_{17}^{-1/7}$, where  $\dot{M}_{17}$
is the rate of mass capture by the neutron star from the stellar
wind of the normal companion expressed in units of $10^{17}\,{\rm
g\,s^{-1}}$.

In the frame of the approach presented in this paper, we associate 
the `missing' flare phenomenon at the orbital phase $\phi=0$ with the
spherical geometry of the accretion flow. The reasons for this can be
high expansion velocity of the Be star envelope and/or a spontaneous
decrease of the parameter $\xi$ slightly below its average value
due to flip-flop instability of the accretion flow.

Finally, the presented approach predicts the average spin up of the
neutron star during moderate flares of $\sim 3.5 \times 10^{-12}\,{\rm
Hz\,s^{-1}}$.

\begin{acknowledgements}
We would like to thank the referee for useful comments and suggesting
improvements and A.E.~Tarasov and V.M.~Lyuty for helpful discussions.
NRI acknowledge the support of the Follow-up program of the
Alexander von Humboldt Foundation. The work was partly supported by INTAS
under the grant YSF 99-4004, RFBR under the grant 99-02-16336 and by the
Federal program ``INTEGRATION'' under the grant KO\,232.
\end{acknowledgements}

\end{document}